\documentstyle[psfig,12pt]{article}
 \newcommand \be {\begin{equation}}
\newcommand \ee {\end{equation}}
 \newcommand \ba {\begin{eqnarray}}
\newcommand \ea {\end{eqnarray}}                        
\begin{document}

\title{\bf Hedging large risks reduces the transaction costs}
\author{F. Selmi$^1$, J.P. Bouchaud$^{2,3}$}
\date{{\small $^1$ {\sc cemefg} Universit\'e Paris II,\\
92-96 rue d'Assas, 75 270 Paris {\sc cedex} 06, FRANCE\\
 $^2$ Science \& Finance, 109-111 rue Victor Hugo,
92532
Levallois {\sc cedex}, FRANCE;\\ http://www.science-finance.fr\\
$^3$ Service de Physique de l'\'Etat Condens\'e,
 Centre d'\'etudes de Saclay,\\
Orme des Merisiers,
91191 Gif-sur-Yvette {\sc cedex}, FRANCE\\}\today}
\maketitle

\begin{abstract}
As soon as one accepts to abandon the zero-risk paradigm of Black-Scholes, 
very interesting issues concerning risk control arise because different 
definitions of the risk become unequivalent. Optimal hedges then depend 
on the quantity one wishes to minimize. We show that a definition of the risk 
more sensitive to the extreme events generically leads to a decrease both of
the probability of extreme losses and of the 
sensitivity of the hedge on the price of 
the underlying (the `Gamma'). Therefore, the transaction costs and the 
impact of hedging on the price dynamics of the underlying are reduced.
\end{abstract}

\vfill

It is well known that the perfect Black-Scholes hedge only works in the
ideal case of a continuous time, log-Brownian evolution of the price of the 
underlying. Unfortunately, this model is rather remote from reality: the 
distribution of price changes has `fat tails', which persist even for rather 
long time lags (see, e.g. \cite{Olsen,Stanley,BP}). This makes the whole idea 
of
zero-risk strategies and perfect replication shady. An alternative view was 
proposed in \cite{BS,BP,Sch}, where one accepts from the 
start that the risk associated with option trading is in general non zero, but 
can be {\it minimized} by adopting an appropriate hedging strategy. If the risk is 
defined as the variance of the global
wealth balance, as was proposed in \cite{BS,BP,Sch}, one can obtain a
expression for the optimal hedge that is valid under rather mild assumptions 
on the dynamics of price changes. This optimal strategy allows one to compute 
the `residual' risk, which is in general non zero, and actually rather large 
in practice. For typical one month at the money options, the minimal standard 
deviation of the wealth balance is of the order of a third of the option 
price itself! This more general theory allows one to recover all the 
Black-Scholes results in the special case of Gaussian returns in the 
continuous time limit, in particular the well known `$\Delta$-hedge', which 
states that the optimal strategy is the derivative of the option price with 
respect to the underlying.

However, as soon as the risk is non zero, the various possible definitions of 
`risk' become unequivalent. One can for example define the risk through a 
higher moment of the wealth balance distribution -- for example the fourth 
moment (whereas the variance is the second moment). This is interesting since 
higher moments are more sensitive to extreme events. The minimisation of the 
fourth moment of the distribution therefore allows one to reduce the 
probability of large 
losses, which is indeed often a concern to risk managers. One could 
also consider the `Value-at-Risk' (defined as the loss level with a certain 
small probability) as the relevant measure of large risks, and 
aim at minimizing that quantity: this is a generalisation of the present work 
which is still in progress \cite{FL}. However, our main conclusions remain valid for 
this case as well.

Our results can be summarized as follows: the optimal strategy obtained using 
the fourth moment as a measure of risk varies much less with the moneyness of the 
underlying than both the Black-Scholes $\Delta$-hedge
and the optimal variance hedge. This is very interesting 
because it means that when the price of the underlying changes, the 
corresponding change in the hedge position is reduced. Therefore, the 
transaction costs associated to option hedging decrease
as one attemps to hedge away large risks. 
Our numerical estimates show that this 
reduction is substantial. This result is also important 
for the global stability of markets: it is well known that the hedging 
strategies can feedback on the dynamics of the markets, as happened during 
the crash of October 1987, where the massive use of the Black-Scholes hedge 
(through `Insurance Portfolio' strategies) amplified the drop of the market. Therefore, part of the `fat-tails' observed 
in the dynamics of price changes can be
attributed to this non-linear feedback effect. By reducing the sensitivity of 
the hedge on the price of the underlying, one can also hope to reduce this 
destabilising feedback.

Let us present our mathematical and numerical results in a rather cursory way 
(a more detailed version will be published separately \cite{Selmi}). 
In order to keep the discussion simple, we will assume that the interest rate 
is zero. In this case, the global wealth balance $\Delta W$ associated to the 
writing of a plain vanilla European call option can be written as:
\be
\Delta W = {\cal C} - \max(x_N-x_s,0) + \sum_{i=1}^{N-1} \phi_i(x_i) 
[x_{i+1}-x_i], \qquad N=\frac{T}{\tau}
\ee
where $\cal C$ is the option premium, $x_i$ the price of the underlying at 
time $t=i\tau$, $\phi(x)$ the hedging strategy, $T$ the maturity of the 
option, $x_s$ the strike and $\tau$ the time interval between rehedging. 
Previous studies focused on the risk defined as 
${\cal R}_2=\langle \Delta W^2 \rangle$, while the fair game option premium 
is such that $\langle \Delta W \rangle=0$ (here, $\langle ... \rangle$ means an 
average over the {\it historical} distribution). As stated above, this allows 
one to recover precisely the standard Black-Scholes results if the statistics of price 
returns is Gaussian and one lets $\tau$ tend to $0$ (continuous time limit). 
This is shown in full detail in \cite{BP}. 

\begin{figure}
\centerline{\psfig{figure=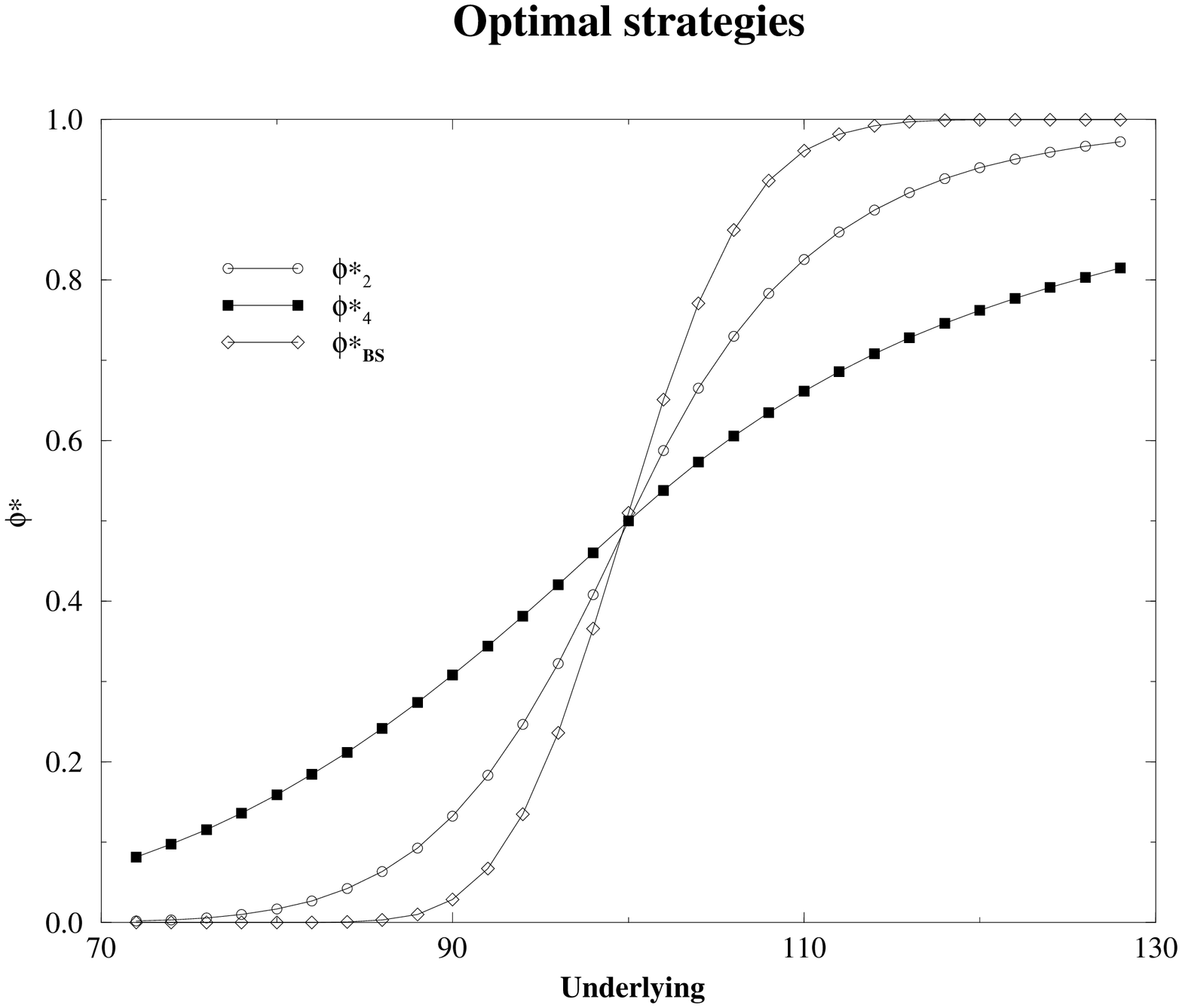,width=8cm,angle=270}}
\caption{Three different strategies: $\phi^*_2$ minimizes the
variance of the wealth balance, $\phi^*_4$ minimizes its fourth moment,
whereas $\phi^*_{BS}$ is the Black-Scholes $\Delta$-hedge. The strike price is $100$,
the maturity equal to one month, the daily volatility is $1\%$ and the terminal price distribution is assumed to be a symmetric exponential, with an excess kurtosis of $3$. The three strategies are equal to $1/2$ at the money. Note
that $\phi_4^*$ varies much less than the other two with moneyness.}
\label{fig1}
\end{figure}

Here, we consider as an alternative 
measure of the risk the quantity ${\cal R}_4=\langle \Delta W^4 \rangle$. The 
corresponding optimal hedge is 
such that the functional derivative of ${\cal R}_4$ with respect to  
$\phi_i(x)$ is zero. This leads to a rather involved cubic equation on 
$\phi_i(x)$ (whereas the minimisation of ${\cal R}_2$ leads to a simple
linear equation on $\phi_i(x)$). Further insight can be gained by first assuming  a time {\it independent} strategy, i.e. $\phi_i(x)\equiv 
\phi_0(x)$. The corresponding cubic equation only depends on the terminal 
price distribution and can then be solved
explicitely, leading to a unique real solution $\phi^*_4$ between $0$ and $1$.
We show in Fig. 1 the evolution of the optimal strategy $\phi^*_4$ as a 
function of the moneyness, in the case where the terminal 
distribution is a symmetric exponential (which is often a good description of 
financial data), for $T=1$ month and a daily volatility
of $1\%$. The corresponding price of the at-the-money call is $2.73$. On the same figure, we have also plotted the Black-Scholes 
$\Delta$-hedge, and the hedge $\phi^*_2$ corresponding to the minimisation of 
${\cal R}_2$. All these strategies vary between zero for deeply 
out of the money options to one for deeply in the money options, which is 
expected. However, as mentioned above, the variation of $\phi^*$ with 
moneyness is much weaker when ${\cal R}_4$ is chosen as the measure of risk. 
For example, for in the money options (resp. out of the money), $\phi^*_4$ is 
smaller (resp. greater) than the Black-Scholes $\Delta$ or than $\phi^*_2$. 
This is because a possible large drop of the stock, which would suddenly 
drive the option out of the money and therefore lead to large losses due to 
the long position on stocks, is better taken into account by considering 
${\cal R}_4$. One can illustrate this result differently by plotting the 
derivative of $\phi^*$ with respect to the price of the stock, which is the 
`Gamma' of the option -- see 
Fig. 2. One sees that in our example the at-the-money Gamma is decreased by a 
factor $3.5$ compared to the Black-Scholes Gamma. The
corresponding average transaction costs for rehedging are therefore also 
expected to decrease 
by the same amount.

\begin{figure}
\centerline{\psfig{figure=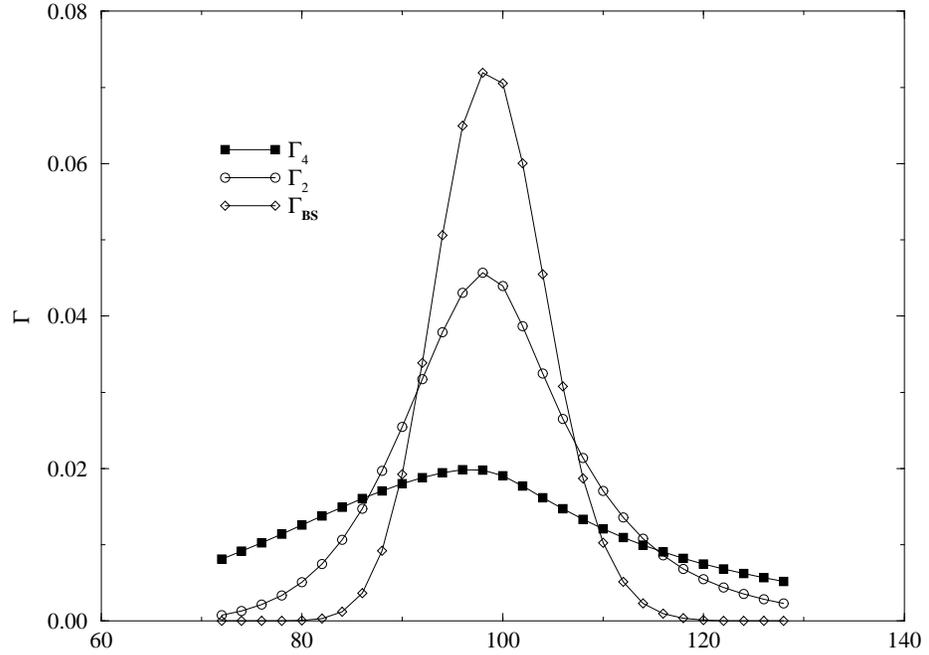,width=8cm,angle=270}}
\caption{The three corresponding `Gamma's', defined as the derivative of the
strategy $\phi^*$ with respect to the price of the underlying. This quantity is
important since the transaction costs for at the money options are 
proportional to $\Gamma(100)$.}
\label{fig2}
\end{figure}

It is interesting to study the full probability distribution function of 
$\Delta W$ for the following three cases: unhedged, 
hedged {\it \`a la} Black-Scholes
or hedged following $\phi^*_4$. Of 
particular interest is the probability $p$ of large losses -- for example, 
the probability of losing a certain amount $|\Delta W^*|$, defined as:
\be
p = \int_{-\infty}^{-|\Delta W^*|} d\Delta W \ P(\Delta W)
\ee
The results for $|\Delta W^*|=10$ (which is four times the option premium) are shown on Fig. 3 for different values of the strike price. One sees that even in the restrictive framework of a purely static
hedge, $\phi^*_4$ allows one to decrease substantially the probability of
large losses. For $x_s=110$, this probability is decreased by a factor $3$ to $4$
as compared to the Black-Scholes hedge! For at-the-money options, since the
static strategies are identical ($\phi^*_{BS}=\phi^*_4=1/2$), one finds no 
difference in $p$. 

\begin{figure}
\centerline{\psfig{figure=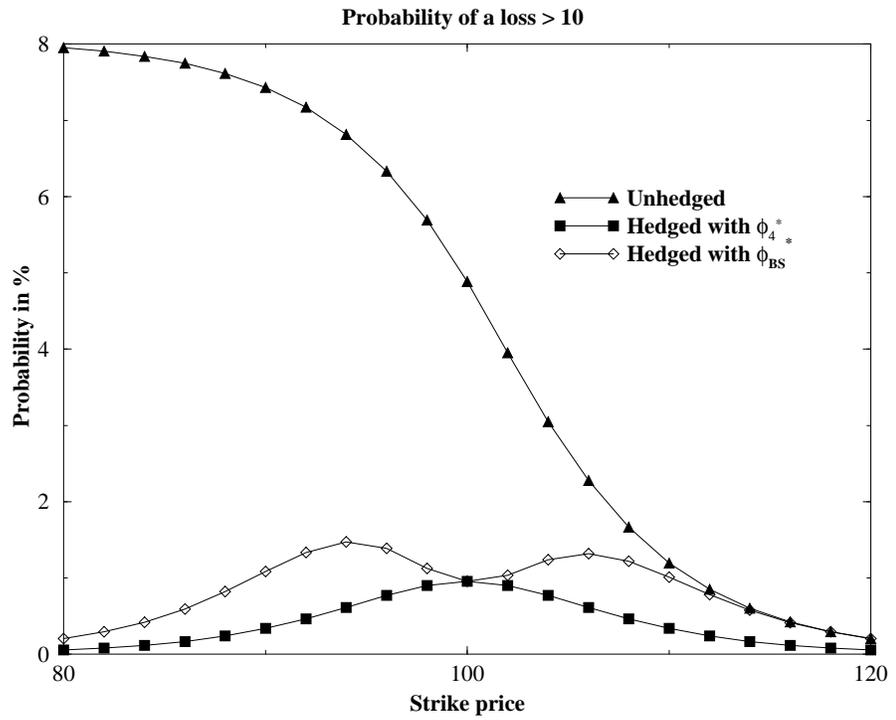,width=8cm,angle=270}}
\caption{The probability $p$ of losing four times the premium, as a function of 
the strike price, for the three different strategies: unhedged, Black-Scholes,
and $\phi_4^*$. Note the substantial decrease of $p$ for out and in the money
options, even in this restrictive case where the strategy is purely static.}
\label{fig3}
\end{figure}

We have up to now considered the simple case of a purely static strategy. In 
the case of the minimisation of ${\cal R}_2$, one can show that the fully 
dynamical hedge can be obtained by a simple time translation of the static 
one, that is, one can compute $\phi_{2i}^*$ by 
again assuming a static hedge, but with an initial time translated from $0$ 
to $t=i\tau$. This can be traced back to the fact that if the price 
increments are uncorrelated (but not necessarily independent), the variance 
of the total wealth balance is the sum of the variances of the 
`instantaneous' wealth balances $\Delta W_i=W_{i+1}-W_i$. This is no longer 
true if one wants to minimise ${\cal R}_4$. However, we have shown for $N=2$ 
that the simple `translated' strategy $\phi^*_4$ is numerically very close to (but 
different from)  
the true optimum. Since we are in the neighbourhood of a quadratic minimum, 
an error of order $\epsilon$ on the strategy will only increase the risk to
order $\epsilon^2$ and is therefore often completely negligible. [Note that a 
similar argument also holds in the case of $\phi_2^*$: even if the 
Black-Scholes $\Delta$ is in general different from $\phi_2^*$, the 
difference is often small and leads to a very small increase of the risk -- 
see the discussion in \cite{BP}]. 

Finally, it is important to note that the optimal $\phi^*_4$
hedge for a book of options on the same underlying is not the simple linear 
superposition of the
optimal hedge for the individual options in the book, whereas this is indeed 
the correct result 
for variance hedging. However, we have found in the case of a book containing two options with different strikes but the same maturity, that the difference between the optimal hedge and the simple linear prescription is 
again numerically very small.

As a conclusion, we hope to have convinced the reader that as soon as one 
accepts to abandon the zero-risk paradigm of Black-Scholes, very interesting 
issues concerning risk control arise because different definitions of the 
risk become unequivalent. [In the Black-Scholes world, the risk is zero, 
whatever the definition of risk !] Therefore, optimal hedges depend on the 
quantity one wishes to minimize. We have shown here that a definition of the 
risk more sensitive to the extreme events generically leads to a decrease of 
the sensitivity of the hedge on the price of 
the underlying (the `Gamma'). Therefore, both the transaction costs and the 
impact of hedging on the price dynamics of the underlying are reduced.
\vskip 0.5cm
\noindent
\underline{Acknowledgements}
We wish to thank M. Potters, A. Matacz, and the students of `Option Math. 
Appli.' of Ecole Centrale Paris for useful discussions.

\end{document}